\documentclass[]{aipproc}

\usepackage{amssym}
\usepackage{graphicx}
\layoutstyle{8x11double}

\begin{document}

\title[Spectra of X-ray Flashes]{Spectral Characteristics of X-ray Flashes 
compared to\\ Gamma-Ray Bursts}

\copyrightyear{2001}

\author{R.~M.~Kippen$^{*}$}{
  address={University of Alabama in Huntsville, Huntsville, AL 35899, USA},
  address={National Space Science \& Technology Center, 320 Sparkman Dr., Huntsville, AL 35805, USA} 
}
\author{P.~M.~Woods$^{**}$}{
  address={Universities Space Research Association, Huntsville, AL 35806, USA},
  address={National Space Science \& Technology Center, 320 Sparkman Dr., Huntsville, AL 35805, USA} 
}
\author{J.~Heise}{
  address={SRON National Institute for Space Research, Sorbonnelaan 2, 3584 CA Utrecht, The Netherlands}
}
\author{J.~J.~M.~in~'t~Zand$^{\S}$}{
  address={Astronomical Institute, Utrecht University, P.O. Box 80 000, 3508 TA Utrecht, The Netherlands}, 
  address={SRON National Institute for Space Research, Sorbonnelaan 2, 3584 CA Utrecht, The Netherlands}
}
\author{M.~S.~Briggs$^{*}$}{
  address={University of Alabama in Huntsville, Huntsville, AL 35899, USA},
  address={National Space Science \& Technology Center, 320 Sparkman Dr., Huntsville, AL 35805, USA}
}
\author{R.~D.~Preece$^{*}$}{
  address={University of Alabama in Huntsville, Huntsville, AL 35899, USA},
  address={National Space Science \& Technology Center, 320 Sparkman Dr., Huntsville, AL 35805, USA}
}

\begin{abstract}
X-ray flashes (XRFs) are a new type of fast transient source observed
with the {\it Beppo\/}SAX Wide Field Cameras (WFC) at a rate of about
four per year.  Apart from their large fraction of 2--26 keV X-rays,
the bulk properties of these events are similar to those of classical
gamma-ray bursts (GRBs).  By investigating the wide-band spectra of
ten events detected in common with WFC and BATSE, we explore the
possibility that XRFs are a low-energy branch of the GRB population.
We find that XRF spectra are similar to those of GRBs, and that their
low peak energies could be an extension of known GRB properties.
\end{abstract}

\maketitle

\section{Introduction}

In the study of prompt emission from gamma-ray bursts (GRBs) one is
overwhelmed by a large amount of observational data with comparatively
little physical understanding.  In this realm, one of the keys to
better understanding lies in the identification of new, or extreme
behavior that differs from that of the bulk ensemble.  One such
enlightening characteristic would be the identification of the lowest
energy bursts, which could help to indicate the limiting form of the
radiation emission process.

Most observed GRBs have energy spectra that exhibit
$\nu\mathcal{F}_{\nu}$ peak power at energies ($E_{\rm peak}$) in the
range $\sim$50--300 keV.  For example, the distribution of
time-averaged $E_{\rm peak}$ for 156 bright bursts measured with {\it
Compton\/}-BATSE \citep{Pre00} is approximately log-normal, with a
centroid of $\sim$175 keV and a width of $\sim$0.5 decade (FWHM).
However, in apparent contradiction to the BATSE results, several
bursts with significant X-ray ($\sim$2--10 keV) emission have been
observed.  In particular, the spectra of several bright bursts
measured with {\it Ginga\/} suggest that either the distribution of
$E_{\rm peak}$ extends below 10 keV, or there is a second (X-ray)
spectral component in some bursts \citep{Str98}.  The later hypothesis
is supported by the fact that 15\% of all BATSE GRBs analyzed show a
low-energy excess above the standard continuum model \citep[]{Pre96}.

The most recent observational clue related to the question of
low-energy bursts is the discovery \citep{Hei01,Hei02} of several
unknown ``X-ray Flashes'' (XRFs; also referred to as ``Fast X-ray
Transients'') using the {\it Beppo\/}SAX Wide Field Cameras (WFC).
These events are distinguished from Galactic transient sources by
their isotropic spatial distribution and short ($\sim$10--100 s)
durations.  Furthermore, they are distinguished from GRBs based on
their non-detection above 40~keV with the {\it Beppo\/}SAX GRB Monitor
--- implying larger X/$\gamma$ ratios than GRBs.  In fact, the
distribution of X/$\gamma$ ratio overlaps considerably with that of
GRBs.  In other respects, such as duration, temporal structure,
spectrum (X-ray) and spectral evolution, XRFs exhibit properties that
are qualitatively similar to the X-ray properties of GRBs.  This
similarity led to the suggestion that the XRFs are in fact ``X-ray
rich'' gamma-ray bursts \citep{Hei01,Hei02}.

\begin{figure}[!!!t]
 \centering
 \includegraphics[height=4.55cm]{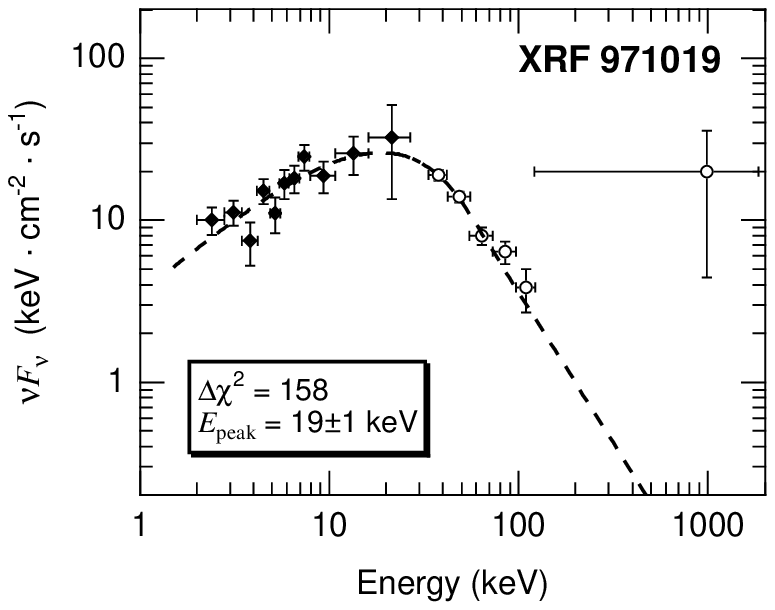}
 \includegraphics[height=4.55cm]{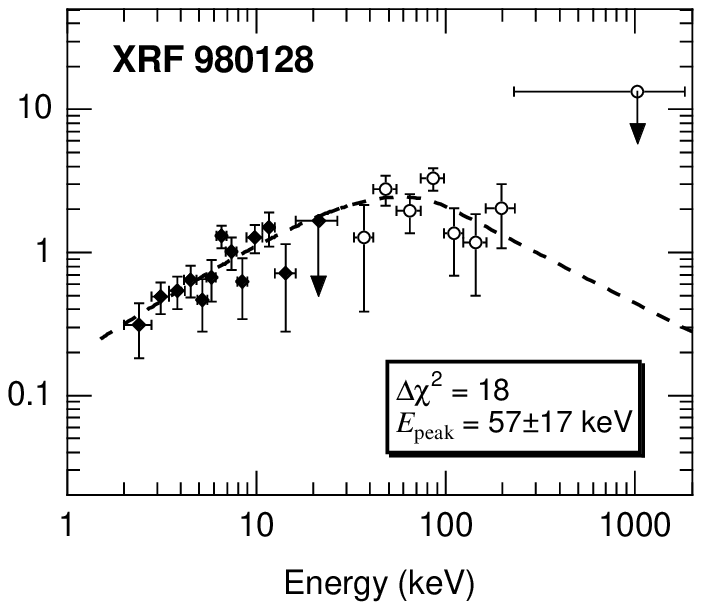}
 \includegraphics[height=4.55cm]{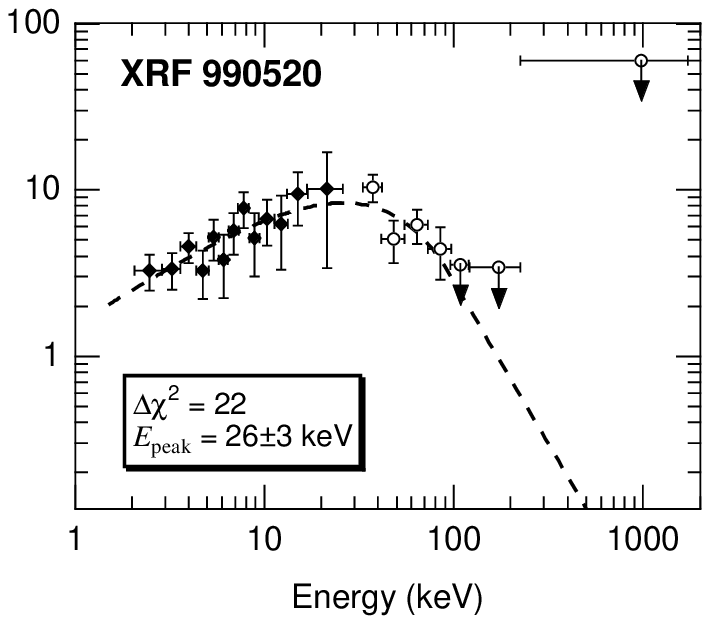}
 \caption{Model-dependent deconvolution of spectral data from 
        WFC (\emph{solid diamonds}) and BATSE (\emph{open circles}) 
        for three X-ray flashes.  The best-fit Band GRB function is
        shown as dashed lines.  Also indicated are the change in 
        chi-squared ($\Delta\chi^2$) from a single power law to
        the Band function and the best-fit values of $E_{\rm peak}$ 
        with 1$\sigma$ errors.}
 \label{fig1}
\end{figure}

To investigate the nature of the WFC X-ray flashes, and their true
relation to GRBs, untriggered BATSE data were searched, and nine of
ten observable flash sources were found to have significant flux above
25~keV (extending in most cases to >100~keV).  These data were used to
directly compare the {\it gamma-ray\/} properties of XRFs to those of
the numerous BATSE GRBs \citep{Kip01}.  This comparison showed that
the flashes are similar in most respects to the long-duration class of
GRBs, except that they are significantly softer (based on gamma-ray
hardness ratios) and weaker (based on gamma-ray peak flux or fluence)
than most long GRBs.  In addition, the XRFs appear to be consistent
with a low-intensity extrapolation of the GRB hardness-intensity (H-I)
correlation --- suggesting that the flashes could indeed represent a
low-energy (or X-ray rich) extension of the GRB population.  If XRFs
are X-ray rich GRBs, they represent a major fraction ($\sim$30\%) of
the full GRB population.

In this paper, we attempt to further quantify the comparison of X-ray
flashes to GRBs by studying wide-band spectral properties.  Only by
combining WFC X-ray data with BATSE gamma-ray data can spectra with
$E_{\rm peak}$ below the gamma-ray regime be parameterized.

\vspace{-0.25cm}
\section{Joint Spectral Analysis}

In the 3.8 years when BATSE and WFC were operating nearly
simultaneously, a total of 36 GRBs and 17 XRFs were detected with WFC.
Due to Earth occultation and data outages, only 18 of the GRB and 10
XRF sources were observable with BATSE.  For these events, we have the
unique opportunity to study the wide-band energy spectra from 2 keV to
$\gtrsim$1 MeV (depending on brightness).  Note that one of the ten
XRF sources is not a significant detection in the BATSE data.  This
event is nonetheless included in the joint spectral analysis because
the BATSE data do constrain the wide-band spectrum.

Joint WFC/BATSE spectral analysis has been successfully applied to
some of the GRBs.  A standard $\chi^2$ fitting technique is used to
compare model spectra, folded through the appropriate instrument
response functions, to the observed (background-subtracted) counting
rates.  For example, the time-averaged spectrum of GRB~990510 was
found to be well described over 3 decades in energy by the
now-standard Band GRB spectral form with $E_{\rm peak} \approx
143$~keV \citep{Bri00}.

Here, the same analysis tools are applied to time-integrated data from
each of the XRF sources.  The WFC data are from source images obtained
by correlating with the WFC coded mask.  The approximate energy range
is 2--26 keV.  The gamma-ray data are from the BATSE Large Area
Detectors (LADs), with 16 energy channels covering the energy range
$\sim$25~keV to 1.8~MeV.  Accumulation time intervals are based on the
entire duration of significant emission in the WFC data.  The spectrum
of each flash was compared to three models: a simple power law (two
free parameters), the ``Comptonized'' model (three free parameters
that describe a power law with a high-energy exponential cutoff), and
Band's GRB model (four free parameters describing two smoothly
connected power laws) \citep[see, e.g.,][]{Pre00}.  The Comptonized
(COMP) and Band models describe a curved spectrum that constrains
$E_{\rm peak}$.

All of the flashes are adequately described by either the COMP or Band
models, which have average $\chi^2/\nu$ values of 0.93 and 0.95,
respectively.  In contrast, the power-law model can be rejected in
most cases, with an average $\chi^2/\nu = 1.57$.  Based on the change
in $\chi^2$ between the single power law and the other models, eight
of the ten XRFs show significant evidence ($\Delta\chi^2 > 4$) for a
curved spectrum resembling that of GRBs.  Examples of spectra for the
three brightest XRFs are shown in Figure~\ref{fig1}.  For these events
the curvature is very significant and the Band spectral parameters are
well-constrained.  For the weaker events the parameters are not as
well constrained, particularly the high-energy spectral index in the
Band function.  This is typical of weak GRBs, where the COMP model is
often used in place of the Band function due to low S/N ratio at high
energies.

\vspace{-0.25cm}
\section{Comparison to GRBs}

When comparing spectral parameters between XRFs and GRBs it is
important to consider biases due to the way the parameters are
measured and how the samples are selected.  Unfortunately, we do not
have large samples of GRBs measured and selected in exactly the same
manner as the XRFs.  Hence, we must make the comparisons with the
data that are available.  The resulting biases cloud the comparison,
as discussed below.

\vspace{-0.25cm}
\subsection{Bright GRBs}

\begin{figure}[t]
 \centering
 \includegraphics[width=6.2cm]{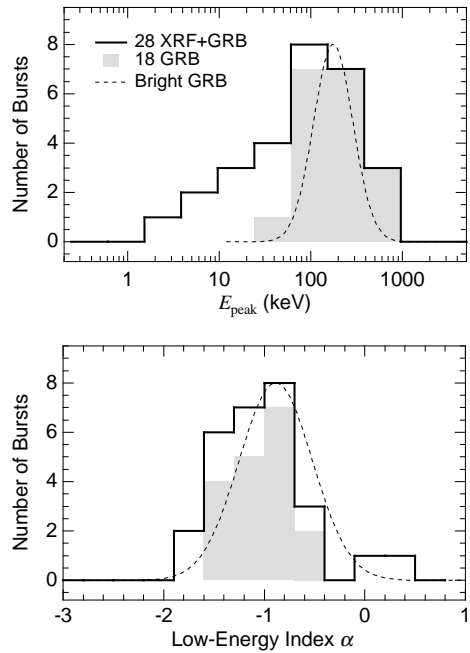}
 \caption{Jointly fit time-averaged spectral parameters of XRFs 
        compared to BATSE-only parameters of GRBs.  Dashed curves 
        are log-normal fits to parameters from 156 bright GRBs 
        \citep[]{Pre00}, normalized to the peaks of the XRF+GRB 
        histograms.}
 \label{fig2}
\end{figure}

The sample of 156 bright, high-fluence BATSE bursts analyzed by Preece
et al.\ \citep[]{Pre00} represents the best current knowledge of
detailed GRB spectral properties.  The high signal-to-noise ratio in
this sample means that time-averaged parameters are well constrained.
In Figure~\ref{fig2}, the distributions of $E_{\rm peak}$ and
low-energy power-law index $\alpha$ for these bursts are compared to
those of the jointly-fit XRFs.  The bright-burst distributions are
represented by best-fit log-normal functions, which provide good
descriptions of the data.  Also included are BATSE-only spectral
parameters for the 18 WFC-selected GRBs observed with BATSE.  These
bursts were fit using BATSE LAD data in the energy range $\sim$25~keV
to 1.8~MeV.  In all cases, the spectral model is the Band GRB
function.  Similar results were obtained using the Comptonized model.

To quantify the comparisons, the K-S test was used to evaluate
statistical differences between the WFC-selected samples and the
log-normal functions for the bright BATSE GRBs.  The results are that
XRFs have significantly lower values of $E_{\rm peak}$ than the bright
BATSE GRBs, with K-S probability $P_{\rm KS} = 1.5\times 10^{-8}$,
while the WFC-selected GRBs have $E_{\rm peak}$'s that are consistent
with those of the bright BATSE bursts.  This latter agreement is not
surprising since most of the WFC GRBs are rather bright.  It does,
however, indicate that (at least for bright bursts) biases due to
different sample selections are small.  Biases due to the fact that
the XRFs were fit over a different energy range (and using different
instrument data) than the GRBs remain unquantified.

The $\alpha$ distributions of WFC XRFs and WFC-selected GRBs are each
statistically consistent with that of bright BATSE bursts, with
$P_{\rm KS} = 0.11$ and 0.32, respectively.  Detailed comparison of
the high-energy power-law index $\beta$ is problematic because it is
not well constrained for many of the XRFs.  However, the mean value
for the four brightest flashes is $\langle\beta\rangle \approx -2.5
\pm 0.5$, which is consistent with the average of $-2.1$ obtained for
bright GRBs.  Thus, excluding the potential measurement biases
mentioned above, it appears that the main outstanding spectral
difference between GRBs and XRFs is that XRFs have lower $E_{\rm
peak}$ values, on average.

\vspace{-0.25cm}
\subsection{Dim GRBs}

Another significant bias that must be considered is the fact that weak
bursts have different spectral properties than bright bursts \citep[see,
e.g.,][]{Nem94,Mal95}.  In particular, there is a strong correlation
between $E_{\rm peak}$ and peak GRB flux (measured in the 50--300 keV
energy range).  Thus, if XRFs are related to GRBs, we expect them to
have lower $E_{\rm peak}$, on average.

For a detailed comparison with weak (and bright) GRBs, we use the
spectral catalog of Mallozzi et al. \citep[]{Mal98}, which contains
model fit parameters for 1,275 BATSE bursts (obtained with LAD data).
Although the catalog contains spectral parameters for several spectral
models, we employ the COMP model, which is more robust for fitting
weak bursts with few high-energy counts.  In practice, the spectral
parameters derived from COMP and Band model fits are similar
\citep[]{Mal98}.

\begin{figure}[!!!t]
 \centering
 \includegraphics[height=4.65cm]{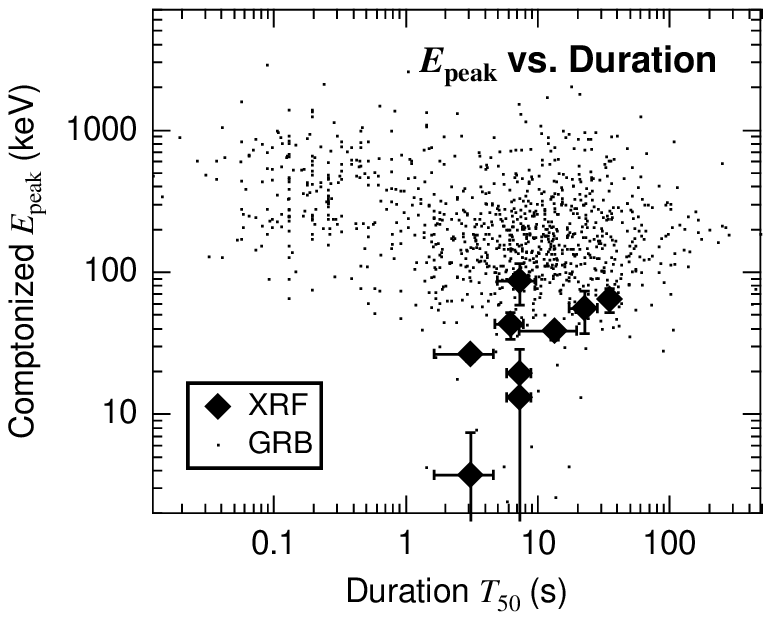}
 \includegraphics[height=4.65cm]{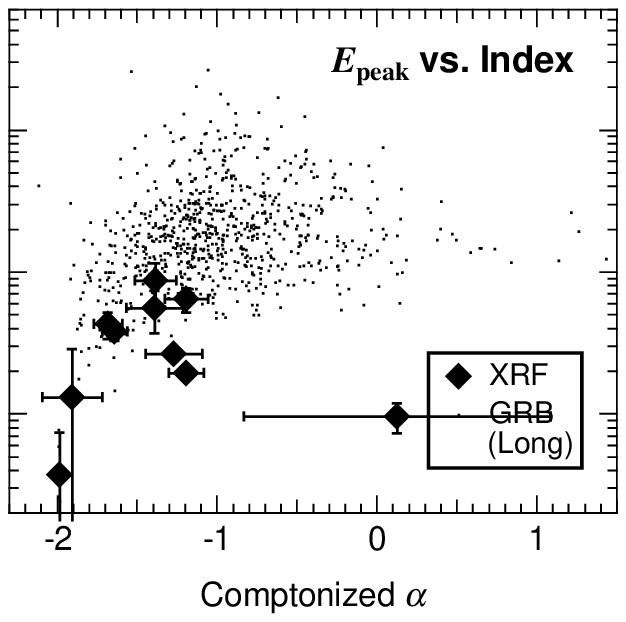}
 \includegraphics[height=4.65cm]{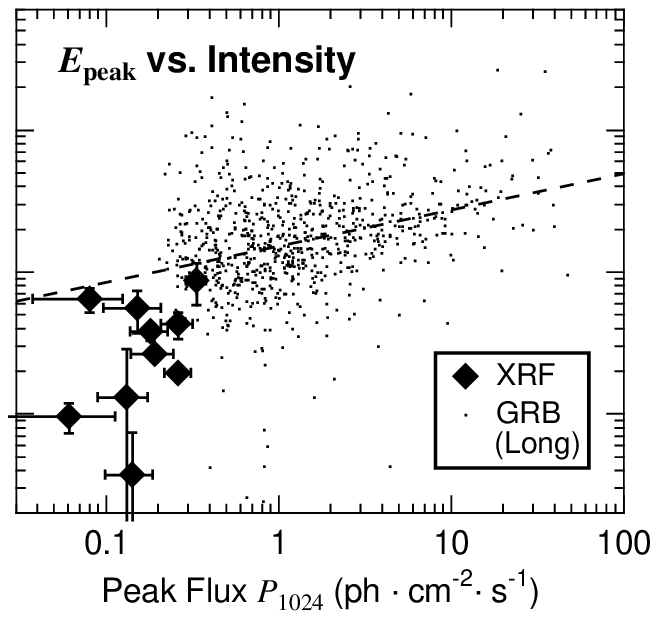}
 \caption{Jointly fit time-averaged spectral parameters of XRFs 
        compared to BATSE-only \citep[]{Mal98} parameters of GRBs.  
        Measurement errors for the GRBs are excluded from the plots 
        for clarity.  The dashed line at right indicates the best-fit
        power law to the GRB distribution.}
 \label{fig3}
\end{figure}

Figure~\ref{fig3} shows distributions of $E_{\rm peak}$ versus
duration ($T_{50}$), power-law index ($\alpha$) and peak flux
($P_{1024}$, evaluated on the 1.024 s timescale in the energy range
50--300 keV) for the XRFs and BATSE GRBs.  Since the XRFs have
(gamma-ray) durations comparable with long GRBs, the Mallozzi et al.\
catalog has been selected for long bursts with $T_{\rm 50} > 1$~s and
that have standard peak fluxes and fluences available.  This results
in a sample of 802 long GRBs that are used for comparison in the two
right-most plots in Figure~\ref{fig3}.

The plot of $E_{\rm peak}$ versus peak flux is particularly
informative since it shows how the XRFs compare with the long GRB H-I
relation.  For this sample of GRBs, the correlation is very
significant, with a Spearman rank order probability of $5.5 \times
10^{-16}$.  At the extremes of the distribution are the bright GRBs
corresponding to the Preece et al.\ sample, with $P_{1024} > 10\ {\rm
ph} \cdot {\rm cm}^{-2} \cdot {\rm s}^{-1}$ and $\langle E_{\rm
peak}\rangle \sim 350$~keV, and the weakest GRBs near the BATSE
trigger threshold with $\langle E_{\rm peak}\rangle \sim 100$~keV.

Statistically, the XRFs are inconsistent even with the weakest 5\% of
GRBs ($P_{1024} < 0.3$), with a two-distribution K-S probability of
$P_{\rm KS} \approx 10^{-5}$.  However, the average XRF flux of
$P_{1024} = 0.16$ is nearly factor of two below that of the weakest
BATSE bursts.

Assuming the H-I correlation extrapolates unchanged below the BATSE
threshold with a simple power law (see Figure \ref{fig3}), the
expected mean and (log-normal) standard deviation of $E_{\rm peak}$ at
the average XRF flux level are 94 keV and 0.25 decades, respectively.
The XRFs are marginally inconsistent with this extrapolation, with
$P_{\rm KS} \approx 10^{-4}$.

\vspace{-0.25cm}
\section{Conclusion}

We have shown that X-ray flashes have significantly curved
time-averaged energy spectra that are quantitatively similar to those
of gamma-ray bursts.  The main difference is that XRFs have peak power
at significantly lower energies than most GRBs.  Combined with our
previous work, this lends further support to the idea that XRFs could
be a low-energy extension of the GRB population.

We also find that, statistically speaking, XRFs are marginally
inconsistent with a power-law extrapolation of the GRB
hardness-intensity distribution.  However, given the considerable
systematic uncertainties in this extrapolation, the marginal
statistical inconsistency is not a strong argument to rule out a
GRB/XRF association.  It is also important to note the two key biases
in this comparison: Firstly, the observed XRFs are purely X-ray
selected, so they probably have lower $E_{\rm peak}$ than average.
Secondly, the XRF spectra were fit using X-ray and gamma-ray data,
whereas the GRBs used only gamma-ray data, so $E_{\rm peak}$ for weak
GRBs could be overestimated.  With these considerations, it is not
unreasonable to conclude that XRFs are indeed low-energy, or X-ray
rich GRBs.

Final conclusion of this problem will undoubtedly require more {\em
wide-band} XRF measurements.  Continued XRF observations with HETE-II
and {\it Beppo\/}SAX may be sufficient, but it will be difficult to
compare these data to known GRBs since they lack high-sensitivity
gamma-ray measurements.

\vspace{-0.25cm}
%%%%%%%%%%%%%%%%%%%%%%%%%%%%%%%%%%%%%%%%%%%%%%%%%%%%%%%%%%%%%%%%%%%%%%%%%%%%%%%%

\end{document}